\documentclass[aps,showpacs,twocolumn]{revtex4}
\usepackage{amsmath}
\usepackage{epsfig}

\begin{document}

\title{Possible existence of a dibaryon candidate $N\Delta$ ($D_{21}$)}

\author{Hongxia Huang$^1$, Xinmei Zhu$^2$, Jialun Ping$^1$\footnote{Corresponding
 author: jlping@njnu.edu.cn}, Fan Wang$^3$}

\affiliation{$^1$Department of Physics and Jiangsu Key Laboratory for Numerical
Simulation of Large Scale Complex Systems, Nanjing Normal University, Nanjing 210023, P. R. China}
\affiliation{$^2$Department of Physics, Yangzhou University, Yangzhou 225009, P. R. China}
\affiliation{$^3$Department of Physics, Nanjing University,Nanjing 210093, P. R. China}

\begin{abstract}
Inspired by the experimental report by WASA-at-COSY Collaboration, we investigate the possibile existence of the
dibaryon candidate $N\Delta$ with quantum numbers $IJ^P=21^+$ ($D_{21}$). The dynamical calculation shows that
we cannot obtain the bound $D_{21}$ state in the models which can obtain the experimental $d^{*}$.
The low-energy scattering phase shifts of the $N-\Delta$ scattering give the same conclusion. Besides, the mass calculation
by using the Gursey-Radicati mass formula and the analysis of the matrix elements of the color magnetic interaction show
that the mass of $D_{21}$ is larger than that of $D_{12}$ ($N\Delta$ with $IJ^P=12^+$), which indicate that it is less possible
for the $D_{21}$ than the $D_{12}$ to form bound state.
\end{abstract}


\maketitle

\setcounter{totalnumber}{5}

\section{\label{sec:introduction}Introduction}
Very recently, an isotensor dibaryon $N\Delta$ with quantum numbers $IJ^P=21^+$ ($D_{21}$) with a mass $M=2140(10)$ MeV
and a width $\Gamma=110(10)$ MeV was reported by WASA-at-COSY~\cite{WASA1}. In their measurements of the quasi-free
$pp \rightarrow pp\pi^{+}\pi^{-}$ reaction by means of $pd$ collisions at $T_{p} = 1.2$ GeV, total and differential
cross sections have been extracted covering the energy region $T_{p} = 1.08-1.36$ GeV, which is the region of
$N^{*}(1440)$ and $\Delta\Delta$ resonance excitations. Calculations describing these excitations by
$t$-channel meson exchange are contradictory with the measured differential cross sections and under-predict
substantially the experimental total cross section. And the new dibaryon $D_{21}$ can be used to overcome
these deficiencies. This state reported by experiment is in good agreement with the prediction of
Dyson and Xuong~\cite{Dyson}. Both mass and width are just slightly smaller than the results of
Faddeev equation calculation by Gal and Garcilazo~\cite{Gal1}. This report invokes our interest to
the dibaryon resonance of the $N\Delta$ system.

The non-strange $S-$wave dibaryon states, labeled as $D_{IS}$ with isospin $I$ and spin $S$, were first proposed
by Dyson and Xuong~\cite{Dyson} in 1964. According to Ref.~\cite{Dyson}, the deuteron $D_{01}$ and $NN$ virtual
state $D_{10}$ were contained within the $\overline{10}$ and $27$ $SU(3)$ multiplets. Besides, four additional
non-strange dibaryon candidates were predicated by the symmetry-based analysis: $\Delta\Delta$ ($D_{03}$) and
$\Delta\Delta$ ($D_{30}$) with mass of $2350$ MeV; $N\Delta$ ($D_{12}$) and $N\Delta$ ($D_{21}$) with mass of $2160$ MeV.
Among these states, the experimental evidence for $D_{03}$ developed in the 1970s by observing a resonance-like
structure in proton polarization from deuteron photodisintegration~\cite{Kamae1}. After that, a lot of theoretical
work have been done supporting the existence of the $\Delta\Delta$ ($D_{03}$ or $d^*$), including the one-boson-exchange model
calculations~\cite{Kamae2,Sato} and quark-model-based calculations~\cite{Mulders1,Mulders2,Oka,Maltman,Goldman,Ping1}.
The positive searching results for $D_{03}$ resonance comes from the exclusive high-statistics measurements of $pn \rightarrow
d\pi^{0}\pi^{0},~\pi^{+}\pi^{-}$ two-pion production reactions by WASA-at-COSY~\cite{WASA2,WASA3,WASA4,WASA5},
and it aroused a new wave of dibaryon studies~\cite{Ping2,Gal2,Bashkanov,Dong}. For $\Delta\Delta$ ($D_{30}$)
state, many former work showed that its mass was higher than that of $\Delta\Delta$ ($D_{03}$), but it was still below the threshold~\cite{Oka,Maltman,Pang,Cvetic,Valcarce,Li}.

For the $N\Delta$ system, the early analyses of $p-p$ and $n-p$ scattering data by Arndt $et~al$ provided
the evidence for the existence of $D_{12}$ in the $I=1$ $^{1}D_{2}$ and $^{3}F_{3}$ nucleon-nucleon
channels~\cite{Arndt}. Some quark models calculation found that $D_{12}$ was almost $200$ MeV too
high~\cite{Mulders1,Mulders2}, but the subsequent chiral quark model calculations showed the mass of it was
about $2170$ MeV, slightly below the threshold of $N\Delta$~\cite{Valcarce}. The recent Faddeev equation
calculation also supported the existence of $N\Delta$ ($D_{12}$)~\cite{Gal1,Gal2}. Moreover,
in Ref.~\cite{Gal2}, another $N\Delta$ state $D_{21}$ was also found slightly below threshold by solving
$\pi NN$ Faddeev equations. However, this $N\Delta$($D_{21}$) was unbound in the early chiral quark model
calculation~\cite{Valcarce}.

In our previous work of dibaryon system, we have showed that the $d^*$ ($D_{03}$) was a tightly bound six-quark
system rather than a loosely bound nucleus-like system of two $\Delta$~\cite{Goldman,Ping2,QDCSM0,QDCSM1,QDCSM2}.
The $\Delta\Delta$ ($D_{30}$) was another dibaryon candidate with smaller binding energy and larger width~\cite{Huang1}.
The $N\Delta$ ($D_{12}$) state could be a resonance state in the $NN$ $D-$wave scattering process only in one
quark model calculation, which gave a lower mass of $d^{*}$, while in other quark model calculations it was
unbound~\cite{Ping2}. Even though a resonance appeared only in one quark model calculation, the mass was very close to
$N\Delta$ threshold. Moreover, the large $\Delta$ decay width when included would cause the state to straddle the
$N\Delta$ threshold in Ref.~\cite{Ping2}. We therefore considered a $N\Delta$ ($D_{12}$) resonance near the $N\Delta$
threshold to be possible in quark model calculations in Ref.~\cite{Ping2}. The $N\Delta$ ($D_{21}$) state, which has
a mirrored quantum numbers for spin and isospin with $D_{12}$, was unbound in our initial calculation~\cite{QDCSM1}.
This situation calls for a more quantitative study of $N\Delta$ ($D_{21}$).

In the various methods of investigating the baryon-baryon interaction, QCD-inspired quark models are still
the main approach, because the direct use of quantum chromodynamics (QCD) in nucleon-nucleon interaction
is still out of reach of the present techniques, although the lattice QCD has made considerable progress
recently~\cite{LQCD}. In our previous study of the dibaryon system, two quark models with different
intermediate-range attraction mechanisms were used: one is the chiral quark model (ChQM)~\cite{Valcarce},
in which the $\sigma$ meson is indispensable to provide the intermediate-range attraction; the other is the
quark delocalization color screening model (QDCSM)~\cite{QDCSM0}, in which the intermediate-range attraction
is achieved by the quark delocalization, and the color screening is needed to make the quark delocalization
feasible and it might be an effective description of the hidden color channel coupling~\cite{Huang2}.
Both QDCSM and ChQM give a good description of the $NN$ scattering phase shifts and the properties of
deuteron despite the different mechanisms used in models~\cite{Chen}. Besides, both models give
$d^*$ ($D_{03}$) resonances reasonable well. Therefore, we use these two models to study the existence of
$N\Delta$ ($D_{21}$) resonance in this work. The hidden color channels are added to the ChQM to check their
effect in the $N\Delta$ system.

The structure of this paper is as follows. A brief introduction
of two quark models is given in Sec. II. Section III is devoted
to the numerical results and discussions. The last section is a
summary.

\section{Two quark models}
In this work, we use two constituent quark models: ChQM and QDCSM, which have been used in our previous
work to study the dibaryon system~\cite{Ping2}. The Salamanca version of ChQM is chosen as the representative of
the chiral quark models. It has been successfully applied to hadron spectroscopy and $NN$ interaction.
The details of two models can be found in Ref.~\cite{Valcarce,QDCSM0,QDCSM1,Ping2}. In the following, only the
Hamiltonians and parameters of two models are given.

\subsection{Chiral quark  model}
The ChQM Hamiltonian in the nonstrange dibaryon system is
\begin{widetext}
\begin{eqnarray}
H &=& \sum_{i=1}^6 \left(m_i+\frac{p_i^2}{2m_i}\right) -T_c
+\sum_{i<j} \left[
V^{G}(r_{ij})+V^{\pi}(r_{ij})+V^{\sigma}(r_{ij})+V^{C}(r_{ij})
\right],
 \nonumber \\
V^{G}(r_{ij})&=& \frac{1}{4}\alpha_s {\mathbf \lambda}_i \cdot
{\mathbf \lambda}_j
\left[\frac{1}{r_{ij}}-\frac{\pi}{m_q^2}\left(1+\frac{2}{3}
{\mathbf \sigma}_i\cdot {\mathbf\sigma}_j \right)
\delta(r_{ij})-\frac{3}{4m_q^2r^3_{ij}}S_{ij}\right]+V^{G,LS}_{ij},
\nonumber \\
V^{G,LS}_{ij} & = & -\frac{\alpha_s}{4}{\mathbf \lambda}_i
\cdot{\mathbf \lambda}_j
\frac{1}{8m_q^2}\frac{3}{r_{ij}^3}[{\mathbf r}_{ij} \times
({\mathbf p}_i-{\mathbf p}_j)] \cdot({\mathbf \sigma}_i+{\mathbf
\sigma}_j),
\nonumber \\
V^{\pi}(r_{ij})&=& \frac{1}{3}\alpha_{ch}
\frac{\Lambda^2}{\Lambda^2-m_{\pi}^2}m_\pi \left\{ \left[ Y(m_\pi
r_{ij})- \frac{\Lambda^3}{m_{\pi}^3}Y(\Lambda r_{ij}) \right]
{\mathbf \sigma}_i \cdot{\mathbf \sigma}_j \right.\nonumber \\
&& \left. +\left[ H(m_\pi r_{ij})-\frac{\Lambda^3}{m_\pi^3}
H(\Lambda r_{ij})\right] S_{ij} \right\} {\mathbf \tau}_i \cdot {\mathbf \tau}_j,  \\
V^{\sigma}(r_{ij})&=& -\alpha_{ch} \frac{4m_u^2}{m_\pi^2}
\frac{\Lambda^2}{\Lambda^2-m_{\sigma}^2}m_\sigma \left[ Y(m_\sigma
r_{ij})-\frac{\Lambda}{m_\sigma}Y(\Lambda r_{ij})
\right]+V^{\sigma,LS}_{ij}, ~~~~
 \nonumber \\
V^{\sigma,LS}_{ij} & = & -\frac{\alpha_{ch}}{2m_{\pi}^2}
\frac{\Lambda^2}{\Lambda^2-m_{\sigma}^2}m^3_{\sigma} \left[
G(m_\sigma r_{ij})- \frac{\Lambda^3}{m_{\sigma}^3}G(\Lambda
r_{ij}) \right] [{\mathbf r}_{ij} \times ({\mathbf p}_i-{\mathbf
p}_j)] \cdot({\mathbf \sigma}_i+{\mathbf \sigma}_j),
\nonumber \\
V^{C}(r_{ij})&=& -a_c {\mathbf \lambda}_i \cdot {\mathbf
\lambda}_j (r^2_{ij}+V_0)+V^{C,LS}_{ij}, \nonumber
\\
V^{C,LS}_{ij} & = & -a_c {\mathbf \lambda}_i \cdot{\mathbf
\lambda}_j
\frac{1}{8m_q^2}\frac{1}{r_{ij}}\frac{dV^c}{dr_{ij}}[{\mathbf
r}_{ij} \times ({\mathbf p}_i-{\mathbf p}_j)] \cdot({\mathbf
\sigma}_i+{\mathbf \sigma}_j),~~~~~~ V^{c}=r^{2}_{ij},
\nonumber \\
S_{ij} & = &  \frac{{\mathbf (\sigma}_i \cdot {\mathbf r}_{ij})
({\mathbf \sigma}_j \cdot {\mathbf
r}_{ij})}{r_{ij}^2}-\frac{1}{3}~{\mathbf \sigma}_i \cdot {\mathbf
\sigma}_j. \nonumber
\end{eqnarray}
\end{widetext}
Where $S_{ij}$ is quark tensor operator, $Y(x)$, $H(x)$ and $G(x)$ are standard Yukawa functions,
$T_c$ is the kinetic energy of the center of mass. All other symbols have their usual meanings.

\subsection{Quark delocalization color screening model}
The QDCSM and its extension were discussed in detail in Ref.\cite{QDCSM0,QDCSM1}. Its Hamiltonian has
the same form as Eq.(1) but without $\sigma$ meson exchange. Besides, a phenomenological color screening
confinement potential is used in QDCSM.
\begin{eqnarray}
V^{C}(r_{ij})&=& -a_c {\mathbf \lambda}_i \cdot {\mathbf
\lambda}_j [f(r_{ij})+V_0]+V^{C,LS}_{ij}, \nonumber
\\
 f(r_{ij}) & = &  \left\{ \begin{array}{ll}
 r_{ij}^2 &
 \qquad \mbox{if }i,j\mbox{ occur in the same } \\
 & \qquad \mbox{baryon orbit}, \\
 \frac{1 - e^{-\mu r_{ij}^2} }{\mu} & \qquad
 \mbox{if }i,j\mbox{ occur in different} \\
 & \qquad \mbox{baryon orbits}.
 \end{array} \right.
\end{eqnarray}
Here, $\mu$ is the color screening constant to be determined by fitting the deuteron mass in this model.
The quark delocalization in QDCSM is realized by replacing the left- (right-) centered single Gaussian functions,
the single-particle orbital wave function in the usual quark cluster model,
\begin{eqnarray}
\phi_{\alpha}(\vec{S}_i) & = & \left( \frac{1}{\pi b^2}
\right)^{3/4}
   e^{-\frac{1}{2b^2} (\vec{r}_{\alpha} - \vec{S}_i/2)^2}  \\
\phi_{\beta}(-\vec{S}_i) & = & \left( \frac{1}{\pi b^2}
\right)^{3/4}
   e^{-\frac{1}{2b^2} (\vec{r}_{\beta} + \vec{S}_i/2)^2}.
\end{eqnarray}
with delocalized ones,
\begin{eqnarray}
\psi_{\alpha}(\vec{S}_i ,\epsilon) & = & \left(
\phi_{\alpha}(\vec{S}_i)
+ \epsilon \phi_{\alpha}(-\vec{S}_i)\right) /N(\epsilon), \nonumber \\
\psi_{\beta}(-\vec{S}_i ,\epsilon) & = &
\left(\phi_{\beta}(-\vec{S}_i)
+ \epsilon \phi_{\beta}(\vec{S}_i)\right) /N(\epsilon), \label{1q} \\
N(\epsilon) & = & \sqrt{1+\epsilon^2+2\epsilon e^{-S_i^2/4b^2}}.
\nonumber
\end{eqnarray}
The mixing parameter $\epsilon(\vec{S}_i)$ is not an adjusted one but determined variationally by the
dynamics of the multi-quark system itself. This assumption allows the multi-quark system to choose its
favorable configuration in a larger Hilbert space. So the ansatze for the wave functions (Eq.~\ref{1q})
is a generalization of the usual quark cluster model ones which enlarges the variational space for the
variational calculation. It has been used to explain the cross-over transition between hadron
phase and quark-gluon plasma phase~\cite{liu}.

Since both of these two models give good descriptions of the deuteron, the nucleon-nucleon scattering
phase shifts, and the dibaryon resonance $d^{*}$ in our previous work~\cite{Ping2}, the same models and
parameters are used in this work. All parameters are listed in Table \ref{parameters}. Here, the same
values of parameters: $b,~\alpha_s,~\alpha_{ch},~m_u,~m_\pi,~\Lambda$ are used for these two models,
which are labeled as ChQM and QDCSM1 in Table \ref{parameters}. Thus, these two models have exactly
the same contributions from one-gluon-exchange and $\pi$ exchange. The only difference of the two models
comes from the short and intermediate-range part, $\sigma$ exchange for ChQM, quark delocalization and
color screening for QDCSM. By doing this, we can compare the intermediate-range attraction mechanism of
these two models. Moreover, another set of parameters in QDCSM (labeled as QDCSM2) is used to test
the sensitivity of the model parameters.

\begin{table}[ht]
\caption{Parameters of quark models}
\begin{tabular}{lccc}
\hline\hline
 & {\rm ChQM} & ~~~~{\rm QDCSM1} & ~~~~{\rm QDCSM2}     \\
\hline
$m_{u,d}({\rm MeV})$        &  313    &  ~~~~313     &  ~~~313 \\
$b ({\rm fm})$              &  0.518  &  ~~~~0.518   &  ~~~~0.60\\
$a_c({\rm MeV\,fm}^{-2})$   &  46.938 & ~~~~56.75    & ~~~~18.55\\
$V_0({\rm fm}^{2})$         &  -1.297 & ~~~~-1.3598  & ~~~~-0.5279 \\
$\mu ({\rm fm}^{-2})$       &   --     &  ~~~~0.45   &  ~~~~1.0    \\
$\alpha_s$                  &  0.485  &  ~~~~0.485   &  ~~~~0.9955  \\
$m_\pi({\rm MeV})$          &  138    &  ~~~~138     &  ~~~~138   \\
$\alpha_{ch}$               &  0.027  & ~~~~0.027    & ~~~~0.027 \\
$m_\sigma ({\rm MeV})$      &  675    &    ~~~~--    &    ~~~~--      \\
$\Lambda ({\rm fm}^{-1})$   &  4.2    &  ~~~~4.2     &  ~~~~4.2\\
\hline\hline
\end{tabular}
\label{parameters}
\end{table}

\section{The results and discussions}
In this work, we study the possibility of the existence of the dibaryon state $N\Delta$ ($D_{21}$). For the first
step, we can estimate the mass of dibaryon states from the dynamical symmetry calculation method. The mass spectrum
of baryons and dibaryons can be obtained by means of the Gursey and Radicati mass formula~\cite{Gursey}
\begin{widetext}
\begin{eqnarray}
M & = & M_{0}+A\cdot C_{SU^{f}(3)}+B\cdot J(J+1)+C\cdot Y +D\cdot [I(I+1)-\frac{Y^{2}}{4}] \nonumber
\end{eqnarray}
\end{widetext}
where the term $C_{SU^{f}(3)}$ is the Casimir operator of the $SU^{f}(3)$ flavour group, and $J$, $Y$ and
$I$ denote the total spin, hypercharge and isospin, respectively. $M_0$ is the average multiplet energy
and the coefficients $A$, $B$, $C$, and $D$ are the strengths of the various splittings. This formula has
been used to calculate the masses of baryons and pentaquarks~\cite{Bijker1,Bijker2}. Here, we use this formula
to calculate the masses of the ground baryons and non-strange dibaryons. All the parameters $M_0$, $A$, $B$, $C$,
and $D$ are determined by fitting the experimental masses of the baryons and dibaryons. The parameter values
obtained in this work are: $M_{0}=1026.2$ MeV (for baryons), $M_{0}=2091.9$ MeV (for dibaryons), $A=9.4085$ MeV,
$B=46.883$ MeV, $C=-194.7$ MeV, and $D=33.218$ MeV. The masses of ground-state baryons and non-strange dibaryons
is listed in Table \ref{mass1} and \ref{mass2}, from which we can see that all ground baryons can be described
well by using this mass formula. For the non-strange dibaryons, we can obtain the experimental mass of the deuteron
($NN$ ($D_{01}$)) and the $d^{*}$ resonance ($\Delta\Delta$ ($D_{03}$)). Although the mass of $\Delta\Delta$($D_{30}$)
is higher than that of $\Delta\Delta$ ($D_{03}$), it is still under the threshold of two $\Delta$s, which indicates
that the $D_{30}$ is a bound state within this method. All these results are consistent with our
quark model calculations~\cite{Huang1}. Besides, we find that the mass of the $N\Delta$ ($D_{12}$) is only $3$ MeV
lower than its threshold, and the mass of $N\Delta$ ($D_{21}$) is $14$ MeV larger than that of $N\Delta$ ($D_{12}$).
Therefore, we cannot obtain the bound $N\Delta$ ($D_{21}$) state within this method.
\begin{center}
\begin{table}[h]
\caption{The mass of baryons (MeV).}
\begin{tabular}{ccccccccc}\hline\hline
  & ~$N$~~ &  ~~$\Lambda$~~  & ~~$\Sigma$~~  & ~~$\Xi$~~  & ~~$\Delta$~~  & ~~$\Sigma^{*}$~~  & ~~$\Xi^{*}$~~  & ~~$\Omega$~~    \\
$Y$ & 1 & 0  & 0 & -1 & 1 & 0 & -1 & -2  \\
$S$ & 0 & 1  & 1 & 2 & 0 & 1 & 2 & 3  \\
$I$ & 1/2 & 0  & 1 & 1/2 & 3/2 & 1 & 1/2 & 0  \\
$J$ & 1/2 & 1/2  & 1/2 & 1/2 & 3/2 & 3/2 & 3/2 & 3/2  \\
$[f]$ & $[21]$ & $[21]$ & $[21]$ & $[21]$ & $[3]$ & $[3]$ & $[3]$ & $[3]$ \\
$M$  & 940 & 1118  & 1184 & 1329 & 1236 & 1381 & 1526 & 1671 \\
$M_{exp.}$  & 939 & 1116  & 1193 & 1318 & 1232 & 1383 & 1533 & 1672 \\\hline
\end{tabular}
\label{mass1}
\end{table}

\begin{table}[h]
\caption{The mass of non-strange dibaryons (MeV).}
\begin{tabular}{ccccccc}\hline\hline
  & ~~$D_{01}$~~ &  ~~$D_{10}$~~  & ~~$D_{03}$~~  & ~~$D_{30}$~~  & ~~$D_{12}$~~  & ~~$D_{21}$~~      \\
$Y$ & 2 & 2  & 2 & 2 & 2 & 2   \\
$S$ & 0 & 0  & 0 & 0 & 0 & 0   \\
$I$ & 0 & 1  & 0 & 3 & 1 & 2 \\
$J$ & 1 & 0  & 3 & 0 & 2 & 1   \\
$[f]$ & $[33]$ & $[42]$ & $[33]$ & $[6]$ & $[42]$ & $[51]$  \\
$M$  & 1876 & 1883  & 2351 & 2394 & 2168 & 2182  \\
$M_{exp.}$  & 1876 & 1878?  & 2380 & ? & 2148? & 2140?  \\\hline
\end{tabular}
\label{mass2}
\end{table}
\end{center}

Then, we do a dynamical calculation to investigate the existence of the dibaryon state $N\Delta$ ($D_{21}$)
within the quark models introduced in Section II. The resonating group method (RGM), described in more detail
in Ref.~\cite{RGM}, is used to calculate the mass of the dibaryon states. The channels involved in the calculation
are listed in Table~\ref{channels}. Here the baryon symbol is used only to denote the isospin, the superscript
denotes the spin, $2S+1$, and the subscript "8" means color-octet, so $^{2}\Delta_{8}$ means the $I, S = 3/2, 1/2$
color-octet state. The hidden-color channel means that the dibaryon composed of two color-octet baryons.
We do four kinds of calculation in this work. The first one is the single-channel calculation of the $N\Delta$ state,
which is labeled as $sc$; the second one is the the $S-$wave channel-coupling calculation, which is labeled as $cc1$;
the third one is the $cc1$ coupling with the $D-$wave channels, labeled as $cc2$; the last one is $cc2$ coupled with
all hidden-color channels, which is labeled as $cc3$. Note that the $cc3$ is carried out only in ChQM, because
the quark delocalization and color screening used in QDCSM were showed to be an effective description of the hidden-color channel-coupling in our previous work of dibaryons~\cite{Ping2,Huang2}, so we do not need to couple the
hidden-color channels again in QDCSM.

\begin{center}
\begin{table}[h]
\caption{The channels with $IJ^{P}=21^{+}$.}
\begin{tabular}{lcccccc}
\hline \hline
           &  \multicolumn{2}{c}{color-singlet} & \multicolumn{3}{c}{color-octet} \\ \hline
   ~$^{3}S_{1}$~ & ~~$N\Delta$~~ & ~~$\Delta\Delta$~~ & ~~$^{2}\Delta_{8}~^{4}N_{8}$~~ &
   ~~$^{4}N_{8}~^{4}N_{8}$~~ & ~~$^{2}N_{8}~^{4}N_{8}$~~ \\
   ~$^{3}D_{1}$~ & $N\Delta$ & $\Delta\Delta$ & $^{2}\Delta_{8}~^{4}N_{8}$ & $^{4}N_{8}~^{4}N_{8}$ &
     $^{2}N_{8}~^{4}N_{8}$ \\
   ~$^{7}D_{1}$~ & $\Delta\Delta$ & & & &  \\ \hline
  \hline
\end{tabular}
\label{channels}
\end{table}

\begin{table}
\caption{The mass $M$ (in MeV) for the $N\Delta$($D_{21}$) state. The threshold of the $N\Delta$ is 2171 MeV.}
\begin{tabular}{ccccccc}
\hline\hline
         &  ~{\rm ChQM}~ & ~{\rm QDCSM1}~  & ~{\rm QDCSM2}~~  \\
 ~~{$sc$}~~  & $2179.8$ & $2179.8$ & $2176.5$  \\
 {$cc1$}  & $2179.2$ & $2178.0$ & $2173.6$  \\
 {$cc2$}  &  $2177.2$ & $2175.4$ & $2170.1$  \\
 {$cc3$}  &  $2176.6$ & $-$ & $-$  \\   \hline\hline
\end{tabular}
\label{BE}
\end{table}
\end{center}

The mass for the $N\Delta$($D_{21}$) state is shown in Table~\ref{BE}, from which we can see that the mass of
the pure $N\Delta$ ($D_{21}$) state is above the $N\Delta$ threshold in all these quark models (ChQM, QDCSM1, and QDCSM2).
This means that the pure $D_{21}$ is unbound. In various kinds of coupling, this state is still unbound, except in QDCSM2
by coupling the $S-$ and $D-$wave channels ($cc2$). The result is similar to that of $N\Delta$ ($D_{12}$) state in our previous calculation~\cite{Ping2}. In Ref.~\cite{Ping2}, the $D_{12}$ was unbound except in QDCSM with the third set of
parameters~\cite{Ping2}, which is labeled as QDCSM2 in the present work. In our previous work, although all these
three models can give good description to the deuteron and the nucleon-nucleon scattering phase shifts~\cite{Ping2},
only ChQM and QDCSM1 fit the data of $d^{*}$ ($\Delta\Delta$ $D_{03}$) well, while the QDCSM2 give too low mass of
the $d^{*}$ resonance. Therefore, we cannot obtain the bound $D_{21}$ state in the models which can obtain the
experimental $d^{*}$, although the $D_{21}$ state can bound in QDCSM2. To investigate the possibility of the existence
of the $D_{21}$ state, more work in depth is needed.

To understand the above results, the effective potentials between $N$ and $\Delta$ with quantum numbers
$IJ^{P}=21^{+}$ are shown in Fig. 1. The effective potential between two colorless clusters is defined as
$V(s)=E(s)- E(\infty)$, where $E(s)$ is the diagonal matrix element of the Hamiltonian of the system in the
generating coordinate. The results show that the potentials are all attractive in three quark models, and
the attraction in QDCSM2 is the largest one, followed by that in QDCSM1, and the lowest one is obtained from ChQM.
However, the attraction in neither of the quark models is strong enough to form the bound state $D_{21}$,
that is why we cannot obtain the pure bound $D_{21}$ state in three quark model calculations.
\begin{figure}[th]
\epsfxsize=3.3in \epsfbox{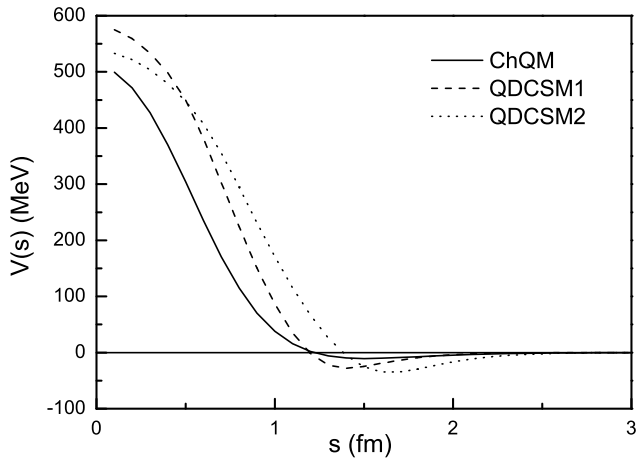} \vspace{-0.1cm}

\caption{The potentials of the $N\Delta$($D_{21}$) state within three quark models. }
\end{figure}

To compare the result to other possible dibaryons, we also calculate the effective potentials of dibaryons
$NN$($D_{01}$), $NN$($D_{10}$), $N\Delta$($D_{12}$), $\Delta\Delta$($D_{03}$), and $\Delta\Delta$ ($D_{30}$),
which are illustrated in Fig. 2. To save space, we only show the results in QDCSM1 here, and the potentials in
other quark models are similar to that in QDCSM1. Although the potentials of all these states are attractive,
only the attractions of $\Delta\Delta$ ($D_{03}$) and $\Delta\Delta$ ($D_{30}$) are strong enough to form bound
states~\cite{Huang1} and the $NN$ ($D_{01}$) state can be bound by coupling $D-$wave channels~\cite{Chen}.
Besides, the attractions of $NN$ ($D_{01}$), $NN$ ($D_{10}$) and $N\Delta$($D_{21}$) are almost same with each other,
while that of the $N\Delta$ ($D_{12}$) state is a little larger. It seems that the possibility of forming a $N\Delta$
($D_{21}$) bound state is smaller than forming a $N\Delta$ ($D_{12}$) state. All these potentials of dibaryons indicate
that the attraction between two decuplet baryons is larger than that between decuplet baryon and octet baryon,
and the attraction between two octet baryons is the smallest one. This regularity has been proposed in our previous work
of dibaryons~\cite{QDCSM1}.
\begin{figure}[th]
\vspace{0.5cm}
\epsfxsize=3.3in \epsfbox{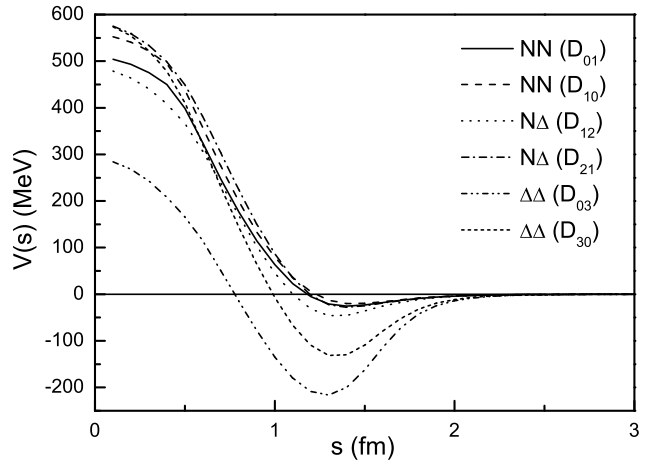} \vspace{-0.2cm}
\caption{The potentials of the dibaryons in QDCSM1. }
\end{figure}

Obviously, the states $N\Delta$ ($D_{12}$) and $N\Delta$ ($D_{21}$) have mirrored quantum numbers of spin and 
isospin with each other, so do the states $NN$ ($D_{01}$) and $NN$ ($D_{10}$), $\Delta\Delta$ ($D_{03}$) and 
$\Delta\Delta$ ($D_{30}$). However, in our previous study of $D_{03}$ and $D_{30}$ states, we found the naive 
expectation of the spin-isospin symmetry was broken by the effective one gluon exchange (OGE) between quarks, 
and the mass of $D_{30}$ state was larger than that of $D_{03}$ state~\cite{Huang1}. So it is interesting to 
study the spin-isospin symmetry of the states $N\Delta$ ($D_{12}$) and $N\Delta$ ($D_{21}$) to see which one 
is more possible to form a bound state. In fact, the mass splitting part in OGE interaction is the color magnetic 
interaction (CMI). It contributes the attraction to the internal energy of an octet baryon, 
$\langle CMI \rangle_{N}=-8C$ (see Eq.(\ref{N}) below, where $C$ is the orbital matrix element, the subscripts 
$A$ and $S$ denote antisymmetric and symmetric), because of the equal attractive and repulsive $qq$ pairs within 
octet baryon. On the contrary there are only repulsive $qq$ pairs within decuplet baryon, 
$\langle CMI \rangle_{\Delta}=8C$ (see Eq.(\ref{D})), which causes the decuplet baryon about $300$ MeV heavier than 
the octet baryon. When two nucleons merge into an orbital totally symmetric color singlet six-quark state, 
the color-spin part of the matrix elements (M.E.) of CMI is $56/3C$ for deuteron $NN$ ($D_{01}$) (see Eq.(\ref{D01p})), 
and $24C$ for $NN$ ($D_{10}$) (see Eq.(\ref{D10p})), which leads to the $NN$($D_{10}$) state massive than $NN$ ($D_{01}$) 
state. At the same time, when two $\Delta$s merge into an orbital totally symmetric color singlet six-quark state, 
the M.E. of CMI is $0C$ for the $d^{*}$ resonance $\Delta\Delta$ ($D_{03}$) (see Eq.(\ref{D03p})), and $32C$ for 
$\Delta\Delta$ ($D_{30}$) (see Eq.(\ref{D30p})), and this causes to the $\Delta\Delta$ ($D_{30}$) state heavier than 
$\Delta\Delta$ ($D_{03}$) state. Similarly, when a nucleon and a $\Delta$ merge into an orbital totally symmetric color 
singlet six-quark state, the M.E. of CMI is $16C$ for the $N\Delta$ ($D_{12}$) (see Eq.(\ref{D12p})), and $80/3C$ for 
$N\Delta$ ($D_{21}$) (see Eq.(\ref{D21p})), and this also indicates that the mass of $N\Delta$($D_{21}$) state is 
larger than that of $N\Delta$ ($D_{12}$) state. All these laws are in complete agreement with the results in 
Table III, which are obtained from the Gursey-Radicati mass formula.
\begin{widetext}
\begin{eqnarray}
\label{N}
\langle CMI \rangle_{N} & = & -3C\langle{\mathbf \lambda}_2 \cdot {\mathbf
\lambda}_3\rangle_{A}[\langle{\mathbf \sigma}_2 \cdot {\mathbf
\sigma}_3\rangle_{A}+\langle{\mathbf \sigma}_2 \cdot {\mathbf
\sigma}_3\rangle_{S}]/2=-8C         \\
\label{D}
\langle CMI \rangle_{\Delta} & = & -3C\langle{\mathbf \lambda}_2 \cdot {\mathbf
\lambda}_3\rangle_{A}\langle{\mathbf \sigma}_2 \cdot {\mathbf
\sigma}_3\rangle_{S}=8C         \\
\langle CMI \rangle_{D_{01}} & = & -C[5\langle{\mathbf \lambda}_5 \cdot {\mathbf
\lambda}_6\rangle_{A}\langle{\mathbf \sigma}_5 \cdot {\mathbf
\sigma}_6\rangle_{A}+13\langle{\mathbf \lambda}_5 \cdot {\mathbf
\lambda}_6\rangle_{A}\langle{\mathbf \sigma}_5 \cdot {\mathbf
\sigma}_6\rangle_{S}+5\langle{\mathbf \lambda}_5 \cdot {\mathbf
\lambda}_6\rangle_{S}\langle{\mathbf \sigma}_5 \cdot {\mathbf
\sigma}_6\rangle_{A}+7\langle{\mathbf \lambda}_5 \cdot {\mathbf
\lambda}_6\rangle_{S}\langle{\mathbf \sigma}_5 \cdot {\mathbf
\sigma}_6\rangle_{S}]/2       \nonumber   \\
\label{D01}
 &=& \frac{8}{3}C         \\
\langle CMI \rangle_{D_{10}} & = & -C[5\langle{\mathbf \lambda}_5 \cdot {\mathbf
\lambda}_6\rangle_{A}\langle{\mathbf \sigma}_5 \cdot {\mathbf
\sigma}_6\rangle_{A}+13\langle{\mathbf \lambda}_5 \cdot {\mathbf
\lambda}_6\rangle_{A}\langle{\mathbf \sigma}_5 \cdot {\mathbf
\sigma}_6\rangle_{S}+7\langle{\mathbf \lambda}_5 \cdot {\mathbf
\lambda}_6\rangle_{S}\langle{\mathbf \sigma}_5 \cdot {\mathbf
\sigma}_6\rangle_{A}+5\langle{\mathbf \lambda}_5 \cdot {\mathbf
\lambda}_6\rangle_{S}\langle{\mathbf \sigma}_5 \cdot {\mathbf
\sigma}_6\rangle_{S}]/2       \nonumber   \\
\label{D10}
 &=& 8C         \\
\label{D03}
 \langle CMI \rangle_{D_{03}} & = & -C[9\langle{\mathbf \lambda}_5 \cdot {\mathbf
\lambda}_6\rangle_{A}\langle{\mathbf \sigma}_5 \cdot {\mathbf
\sigma}_6\rangle_{S}+6\langle{\mathbf \lambda}_5 \cdot {\mathbf
\lambda}_6\rangle_{S}\langle{\mathbf \sigma}_5 \cdot {\mathbf
\sigma}_6\rangle_{S}]=16C         \\
\label{D30}
\langle CMI \rangle_{D_{30}} & = & -C[9\langle{\mathbf \lambda}_5 \cdot {\mathbf
\lambda}_6\rangle_{A}\langle{\mathbf \sigma}_5 \cdot {\mathbf
\sigma}_6\rangle_{S}+6\langle{\mathbf \lambda}_5 \cdot {\mathbf
\lambda}_6\rangle_{S}\langle{\mathbf \sigma}_5 \cdot {\mathbf
\sigma}_6\rangle_{A}]=48C         \\
\langle CMI \rangle_{D_{12}} & = & -C[\langle{\mathbf \lambda}_5 \cdot {\mathbf
\lambda}_6\rangle_{A}\langle{\mathbf \sigma}_5 \cdot {\mathbf
\sigma}_6\rangle_{A}+8\langle{\mathbf \lambda}_5 \cdot {\mathbf
\lambda}_6\rangle_{A}\langle{\mathbf \sigma}_5 \cdot {\mathbf
\sigma}_6\rangle_{S}+2\langle{\mathbf \lambda}_5 \cdot {\mathbf
\lambda}_6\rangle_{S}\langle{\mathbf \sigma}_5 \cdot {\mathbf
\sigma}_6\rangle_{A}+4\langle{\mathbf \lambda}_5 \cdot {\mathbf
\lambda}_6\rangle_{S}\langle{\mathbf \sigma}_5 \cdot {\mathbf
\sigma}_6\rangle_{S}]       \nonumber   \\
\label{D12}
 &=& 16C         \\
 \langle CMI \rangle_{D_{21}} & = & -C[\langle{\mathbf \lambda}_5 \cdot {\mathbf
\lambda}_6\rangle_{A}\langle{\mathbf \sigma}_5 \cdot {\mathbf
\sigma}_6\rangle_{A}+8\langle{\mathbf \lambda}_5 \cdot {\mathbf
\lambda}_6\rangle_{A}\langle{\mathbf \sigma}_5 \cdot {\mathbf
\sigma}_6\rangle_{S}+4\langle{\mathbf \lambda}_5 \cdot {\mathbf
\lambda}_6\rangle_{S}\langle{\mathbf \sigma}_5 \cdot {\mathbf
\sigma}_6\rangle_{A}+2\langle{\mathbf \lambda}_5 \cdot {\mathbf
\lambda}_6\rangle_{S}\langle{\mathbf \sigma}_5 \cdot {\mathbf
\sigma}_6\rangle_{S}]       \nonumber   \\
\label{D21}
 &=& \frac{80}{3}C       \\
\label{D01p}
\langle CMI \rangle_{D_{01}}&-&2\langle CMI \rangle_{N} = \frac{56}{3}C  \\
\label{D10p}
\langle CMI \rangle_{D_{10}}&-&2\langle CMI \rangle_{N} = 24C  \\
\label{D03p}
\langle CMI \rangle_{D_{03}}&-&2\langle CMI \rangle_{\Delta} = 0  \\
\label{D30p}
\langle CMI \rangle_{D_{03}}&-&2\langle CMI \rangle_{\Delta} = 32C  \\
\label{D12p}
\langle CMI \rangle_{D_{12}}&-&(\langle CMI \rangle_{N}+\langle CMI \rangle_{\Delta}) = 16C  \\
\label{D21p}
\langle CMI \rangle_{D_{21}}&-&(\langle CMI \rangle_{N}+\langle CMI \rangle_{\Delta}) = \frac{80}{3}C  \\
\end{eqnarray}
\end{widetext}

Finally, to further check the existence of the bound state $N\Delta$ ($D_{21}$), we can also study the low-energy 
$N-\Delta$ scattering phase shifts. Here we use the well-developed Kohn-Hulthen-Kato (KHK) variational 
method to calculate the scattering phase shifts. The details can be found in Ref.~\cite{RGM}. The phase shifts are 
illustrated in Fig. 3. It is obvious that in QDCSM2, the scattering phase shifts go to $180^{\circ}$ at the 
incident energy $E_{c.m.}\sim 0$ and rapidly decreases as $E_{c.m.}$ increases, which implies the existence of a 
bound state. The results are consistent with that of the bound state calculation shown above. Meanwhile, the slop 
of the low-energy phase shifts (near $E_{c.m.}\sim 0$) in both ChQM and QDCSM1 is opposite to that in QDCSM2, 
which means that the $N\Delta$ ($D_{21}$) is unbound in both ChQM and QDCSM1, and this results is also consistent 
with that of the bound state calculation above.

\begin{figure}
\epsfxsize=3.3in \epsfbox{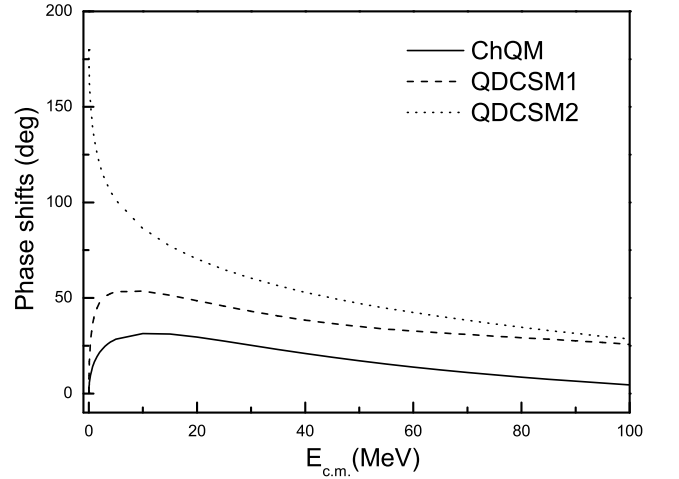} \vspace{-0.2cm}
\caption{The phase shifts the $N\Delta$($D_{21}$) state within three quark models. }
\end{figure}

Then we extract the scattering length $a_{0}$ and the effective range $r_{0}$ from the
low-energy scattering phase shifts by using the formula
\begin{eqnarray}
kcot\delta & = & -\frac{1}{a_{0}}+\frac{1}{2}r_{0}k^{2}+{\cal
O}(k^{4})
\end{eqnarray}
where $k$ is the momentum of relative motion with $k=\sqrt{2\mu E_{c.m.}}$, and $\mu$ is the reduced mass of $N$ 
and $\Delta$. The results are listed in Table~\ref{ar}. From Table~\ref{ar}, we can see that in both ChQM and QDCSM1, 
the scattering lengths $a_{0}$ are negative, while in QDCSM2 the scattering length is positive, which implies the unbound
$N\Delta$ ($D_{21}$) in both ChQM and QDCSM1 and bound $N\Delta$ ($D_{21}$) in QDCSM2.

\begin{center}
\begin{table}
\caption{The scattering length $a_{0}$ and the effective range $r_{0}$ of the $N\Delta$($D_{21}$) state.}
\begin{tabular}{ccccccc}
\hline\hline
         &  ~{\rm ChQM}~ & ~{\rm QDCSM1}~  & ~{\rm QDCSM2}~~  \\
 ~~{$a_{0}~(fm)$}~~  & $-1.9939$ & $-2.5159$ & $6.0756$  \\
 {$r_{0}~(fm)$}  & $2.2974$ & $1.7352$ & $1.5625$  \\
 \hline\hline
\end{tabular}
\label{ar}
\end{table}
\end{center}

\section{Summary}
In the present work, we investigate the possible existence of the dibaryon state $N\Delta$ ($D_{21}$).
The dynamical calculation shows that although the potentials are all attractive in three quark models, 
and the attraction is not strong enough to form the bound state $N\Delta$ ($D_{21}$) in the single channel 
calculation. By various kinds of coupling, this state is still unbound, except in QDCSM2 by coupling the $S-$ 
and $D-$wave channels. But the QDCSM2 gives too large binding energy for the $d^{*}$ ($D_{03}$) resonance. 
Therefore, we cannot obtain the bound $N\Delta$ ($D_{21}$) state in the models which can obtain the 
experimental $d^{*}$. We also study the low-energy scattering phase shifts of the $N\Delta$ ($D_{21}$) state 
and the same conclusion is obtained.

Both the mass calculation by using the Gursey-Radicati mass formula and the analysis of the color-spin part of 
the matrix elements of the color magnetic interaction show that the mass of $NN$ ($D_{10}$) is larger than that 
of $NN$ ($D_{01}$), the mass of $N\Delta$ ($D_{21}$) larger than that of $N\Delta$ ($D_{12}$), and the mass of 
$\Delta\Delta$ ($D_{30}$) larger than that of $\Delta\Delta$ ($D_{03}$). All these results indicate that it is 
less possible for the $N\Delta$ ($D_{21}$) than the $N\Delta$($D_{12}$) to form bound state. 
Besides, the naive expectation of the spin-isospin symmetry is broken by the effective one gluon exchange between 
quarks. The non-strange dibaryon states searching will be another check of this
gluon exchange mechanism and the Goldstone boson exchange model.

In our previous study of $NN$ and $\Delta\Delta$ systems, both ChQM and QDCSM1 can obtain similar results. 
Here, in the work of $N\Delta$ system, the similar results are obtained again. This show once more that the 
$\sigma$-meson exchange used in the chiral quark model can be replaced by quark delocalization and
color screening mechanism.

\section*{Acknowledgment}
This work is supported partly by the National Science Foundation
of China under Contract Nos. 11675080, 11775118 and 11535005, the Natural Science Foundation of
the Jiangsu Higher Education Institutions of China (Grant No. 16KJB140006).

\end{document}